# Highlights

1. For a deep quench, the effect of lower fractions of disorder on segregation kinetics is almost negligible.
2. For a shallow quench and lower fractions of disorder, the scaling functions slightly deviate as the system evolves to form fragmented lamellar stripes.
3. For a high fraction of disorder, lamellar patterns are formed eventually at all the quench depths.
4. For a high fraction of disorder, a crossover in length scale $\phi \sim 1/3 \rightarrow 1/2$ is observed that saturates at late times.
5. Morphologies align in the direction of a higher number of disorder sites on the lattice.

# TOC

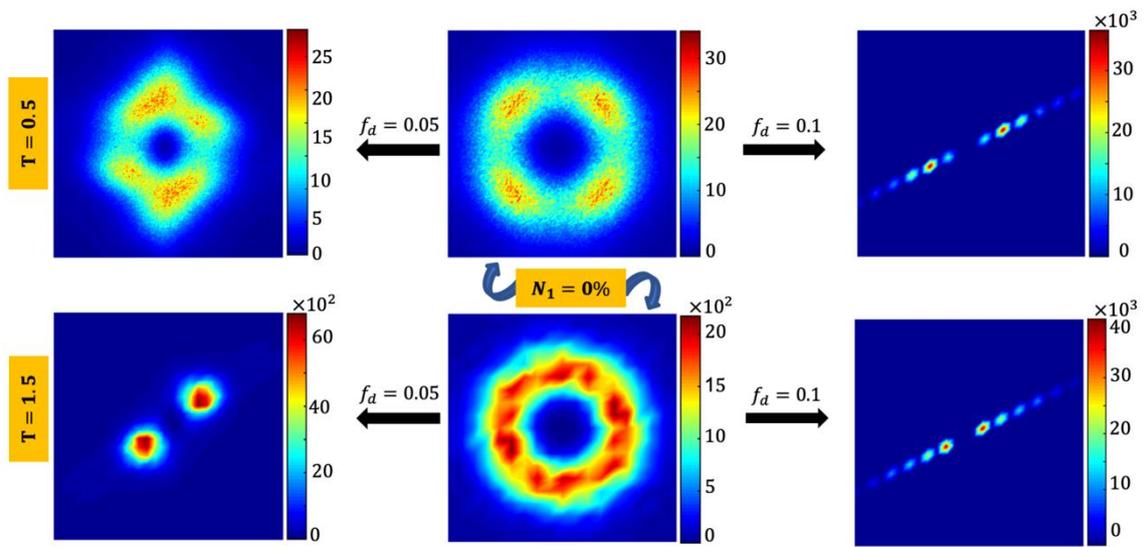

We studied the effect of deep and shallow quench temperatures on phase separation kinetics of a critical binary mixture under the influence of bond disorder introduced in a regular manner. The separation kinetics was modeled by the conserved Kawasaki spin-exchange kinetics of the Ising model on a square kinetic using the Monte Carlo (MC) simulation technique.

# Phase separation kinetics of binary mixture in the influence of bond disorder: Sensitivity to quench temperature

Samiksha Shrivastava and Awaneesh Singh*

Department of Physics, Indian Institute of Technology (BHU), Varanasi-221005, India

**Abstract**

Morphologies in phase separating systems can significantly influence the final properties of materials. We present extensive Monte Carlo (MC) simulation results on the segregation kinetics of the critical binary ($AB$) mixture with a fraction of bond disorder (BD) introduced in a regular manner. We focus on studying the effect of various quench temperatures on the growth kinetics and scaling properties of evolving morphologies. The two-dimensional ($2d$) kinetic Ising system with conserved spin exchange kinetics is used to model the system. We observe that domain morphologies change from their usual interconnected bicontinuous isotropic patterns at zero BD to short strips and lamellar patterns (anisotropy) with increasing BD at shallow quench. The domain evolution remains extremely slow at deep quench and for lower fractions of BD, and thus, morphologies appear very similar; however, we observed lamellar patterns at high BD. The scaling behavior represented by the correlation function and the structure factor changes significantly with quench depths for a higher fraction of BD. In contrast, a tiny deviation from the scaling is observed at a lower fraction of disorder for shallow quenches. The growth law is consistent with the *Lifshitz-Slyozov* (LS) growth law ($\phi \to 1/3$) for shallow quench and at low fractions of BD studied here. At a high fraction of BD, the length scale crossovers gradually from an early time LS growth to the diffusion dynamics ($\phi \to 1/2$) during intermediate times for both deep and shallow quenches. The domain growth freezes ($\phi \to 0$) to a finite size when the system evolves to an equilibrium (stable) lamellar morphology in the asymptotic time limit on the time scale of our simulation. However, no significant changes are observed in the scaling behavior at lower fractions of BD for the deep quench.

**Keywords**: Phase separation, Kawasaki kinetics, critical binary mixture, domain growth, bond disorder.

*Author for the correspondence: awaneesh.phy@iitbhu.ac.in; awaneesh11@gmail.com

## 1. Introduction

A binary ($AB$) mixture is homogeneously mixed (disordered state) at high temperatures ($T > T_C$) where $T_C = 2.269 \, J/k_B$ implies the critical temperature at which a $2d$ Ising system changes its physical behavior.[1–3] When a homogeneous binary ($AB$) mixture is quenched below the critical temperature ($T < T_C$), it becomes thermodynamically unstable due to small inhomogeneities initiated within the system. The phase separation in this *far-from-equilibrium* system begins either by the spinodal decomposition (SD)[1–5] or by the nucleation and growth (NG)[2–5] enriched in either component.

The kinetics of phase separation has been a well-studied subject of interest for decades, focusing on the coarsening of binary mixtures utilizing experimental[6–8], analytical[2,9], and simulation methods[10–14]. Nonetheless, this is still a highly active and robust area of research.[15–21] To specify the coarsening morphologies, one typically calculates the following two crucial physical quantities of practical importance; (i) the two-point equal-time correlation function $C(\vec{r}, t)$ where $\vec{r} = \vec{r}_1 - \vec{r}_2$ and its Fourier transforms, the structure factor $S(\vec{k}, t)$ where $\vec{k}$ represents the wave vector[1,2]; (ii) the domain growth law: time evolution of characteristic domain size, which depends on a few common system properties such as conservation laws, hydrodynamic velocity field[22,23], and the presence of quenched or annealed disorder[16,24–28].

The domain growth law of a pure isotropic system typically follows power-law behavior: $\ell(t) \sim t^\phi$ at late times where $\phi$ denotes the growth exponent.[1,2] The value of $\phi$ depends on the mechanism that drives the coarsening. For the diffusion-driven segregation of conserved $AB$ mixture, the rate of domain evolution scales as $\dot{\ell}(t) \sim |\vec{\nabla}\mu| \sim \sigma/\ell(t)^2$ provides $\ell(t) \sim t^{1/3}$; this is commonly known as the Lifshitz-Slyozov (LS) growth law with growth exponent: $\phi = 1/3$.[1–3,29] Here, $\dot{\ell}(t)$ denotes the interface velocity, $\mu$ is the chemical potential, and $\sigma$ is the interfacial tension between the phases. Nonetheless, with hydrodynamic effects in the system, various growth regimes appear depending on the dimensionality and the other system parameters.[22,23,30–32]

Typically, it is hard to find a pure and isotropic experimental system as it always comprises some impurities (annealed or quenched). In this context, a few essential studies (analytical and numerical) are performed on the $2d$ Ising model with the quenched disorder.[9–12,24–26] The quench-disorder is considered as an immobile impurity in the system; it is introduced in a pure Ising system in the following way: (i) the random-bond Ising model (RBIM)[12,24–26] using $nn$ spin-exchange kinetics, and (ii) the random-field Ising model (RFIM).[33–36] The disordered sites trap domain boundaries and result in slower domain growth. A critical study of RBIM for the non-conserved system is done by Huse and Henley (HH).[28] They argued that the energy barrier ($E_b$), which traps evolving domain boundaries, follows power-law dependence on domain size ($\ell$) as $E_b(\ell) \sim \varepsilon \ell^\psi$, where $\varepsilon$ denotes the disorder/impurity strength.

The barrier exponent $\psi = \chi/(2 - \zeta)$ where the roughening ($\zeta$) and pinning exponent ($\chi$) are related as $\chi = 2\zeta + d - 3$; here, $d$ is the system dimensionality. The characteristic length scale turns out to follow a logarithmic behavior: $\ell(t) \sim (\ln t)^\phi$. Subsequently, many experimental[6–8] and simulation[9–14,26,27] works were performed to assess the HH proposal without explicitly confirming the universal logarithmic growth law in the asymptotic regime. Nevertheless, a power-law growth with the variable exponent is suggested instead.

The RBIM was further revisited by Paul, Puri, and Rieger (PPR) in detail using Monte Carlo (MC) simulations of kinetic Ising models with non-conserved spin-flip kinetics and conserved spin-exchange kinetics.[24,25] PPR proposed that the energy barrier for trapping domain boundaries follows logarithmic dependence on the domain size: $E_b(\ell) \sim \varepsilon \ln(1 + \ell)$ instead of a power law.[24,25] In contrast to HH observation, PPR showed a power-law dependence: $\ell(t) \sim A(\varepsilon, T) t^{\theta(\varepsilon,T)}$ on the average domain size where growth exponent $\theta(\varepsilon, T) = 1/(3 + \varepsilon T^{-1})$ depends on the quench depth ($T$) and impurities ($\varepsilon$) in the system; these results were further confirmed experimentally.[6–8] Overall, the energy barrier becomes negligible due to the small average domain size at early times; hence, the system evolves like a pure system. However, after a crossover length scale at late times, disorder traps become effective due to a higher energy barrier.[28] Therefore, domain coarsening occurs via thermal activation ($\sim k_B T$) over the corresponding energy barrier ($E_b(\ell)$); $k_B$ denotes the Boltzmann constant. Thus, thermal fluctuations drive the domain growth at late times[23,24], contrasting with the pure case, where thermal fluctuations are irrelevant.

Nonetheless, HH and PPR have introduced the quenched disorder by uniformly varying the spin coupling strength between zero and one for all the nearest neighbor ($nn$) sites.[24,25,28] Moreover, a couple of recent studies[16,21] demonstrate the effect of quenched disorder on the $2d$ Ising system where BD is introduced in the system in the (i) random manner and (ii) regularly manner[16]. Authors have studied the evolution kinetics only for a moderate quench at $T = 1.0$ and up to an intermediate time $t = 1.6 \times 10^6$ MCS. The nearest neighbor spin-spin coupling strength is set to one involving pure sites and zero with any of the disordered site. The insights from the previous works[16] lead us to study the effect of shallow ($T = 1.5$) and deep ($T = 0.5$) quenches on segregation kinetics for different fractions of BD introduced in a regular manner. Herein, we let the system evolve for a much longer time $t = 4.0 \times 10^6$ MCS than in the previous studies[16,21] to expound the effect of shallow or deep quench depth more precisely on the final equilibrium morphology at a different fraction of BD. Nevertheless, we also reconsider the moderate quench ($T = 1.0$) case, which is unexplored in the asymptotic time limit ($t = 4.0 \times 10^6$ MCS), particularly at a lower fraction of BD. Overall, our emphasis is on understanding the effect of deep and shallow quenches on domain growth and dynamic universality in the system with BD introduced in a regular manner, where theoretical calculations are usually much more complex.

We organize this paper as follows. First, in Sec. 2, we briefly explain the numerical methodology used to simulate the system. Then, Sec. 3 presents the results and discussions for different temperatures and the fraction of BD. Finally, in Sec. 4, we conclude this paper with a summary of our results.

## 2. Simulation model and implementation

***Basic setup with disordered sites:*** MC simulation method is utilized to study the effect of BD in the $2d$ Ising system. The Ising Hamiltonian is as follows:

$$H = -\sum_{\langle ij \rangle} J_{ij} S_i S_j - \sum_{i=1}^{N} h_i S_i, \quad S_i = \pm 1 \qquad (1)$$

where $S_i$ denotes the spin at $i^{th}$-lattice site. For a binary ($AB$) mixture, it can take the value, $S_i = +1$ (up spin state) when site $i$ is occupied by an $A$-type atom or $S_i = -1$ (down spin state) a $B$-type atom. The parameter, $J_{ij}$ represents the spin-exchange coupling strength for nearest-neighbor ($nn$) spin pairs, and $\langle ij \rangle$ indicates the sum over those spins. In general, $J_{ij} > 0$ is set for a ferromagnetic system, and $J_{ij} < 0$ for an antiferromagnetic system. However, a system unified with ferromagnetic and antiferromagnetic exchange coupling is usually relevant to spin glasses. The external field in our simulation is set to zero ($h_i = 0$); nevertheless, $h_i \neq 0$ for $J_{ij} > 0$ is known as RFIM[33–36], one of the most straightforward Ising systems with quenched disorder. We set, $J_{ij} = 1$ for a pure system; the corresponding critical temperature $T_c \approx 2.269 J_{ij}/k_B$ which is obtained from Onsager's exact solution of the $2d$ Ising model on a square lattice with $nn$ interactions.[37]

In our simulation, we choose a square lattice of size $N = L_x \times L_y$ with periodic boundary conditions in both directions where $L_x = L_y = 512$. The bond disorder is introduced via the exchange coupling parameter: $J_{ij} = 1 - \varepsilon_{ij}$, where $\varepsilon_{ij}$ quantifies the degree of disorder.[16] For simplicity, we consider two limiting cases: (i) $\varepsilon_{ij} = 0$, corresponds to a pair of pure $nn$ sites, and (ii) $\varepsilon_{ij} = 1$ corresponds to a disordered site in a $nn$ spin pair (system's impurity).[16] Thus, $J_{ij} = 1$ resembles connecting any two pure neighboring sites, and $J_{ij} = 0$ when any of the two neighboring sites is impure (disordered). The relevant configuration of up and down spins, tagged on pure and disordered sites, are displayed schematically in Figs. 1(a)-(c). Note that in PPR's RBIM[24,25] study, a uniform distribution of $J_{ij} \in (0,1)$ was considered. Thus, our study is a limiting case of PPR's study regarding the value of exchange coupling strength. To introduce the disorder, we select a fraction of sites in a regular manner, as displayed in Fig. 1(a) schematically with red circles. For every lattice index from $1 \cdots L_y$ (in $y$-direction), we traced all the indexes from $1 \cdots L_x$ (in $x$-direction) and in the process, every $m^{th}$ site is tagged as the disorder site. Thus, the total number of disordered sites in a system is, $N_d = f_d N$ where $f_d = 1/m$ represents the fraction of disordered sites. In Figs. 1(d)-(e), we displayed a section of disordered sites with $f_d = 0.02$, 0.05, and 0.1, respectively, on a $2d$ square lattice of size $N = L^2$ where $L = 512$.

The Ising model itself does not have any intrinsic dynamic. We place the system in contact with a heat bath to introduce the stochastic dynamics.[1–3] Thus, the resultant dynamical model is referred to as a *kinetic Ising model*.[1] We exploit the Kawasaki spin-exchange (conserved) kinetics as an appropriate stochastic kinetics to model the phase separation in a binary ($AB$) mixture. In this conserved kinetics, we randomly select two $nn$ sites with opposite spins to exchange them ($S_i \leftrightarrow S_j$). The energy change for the spin-exchange to take place is given by[38]

$$\Delta H_{ij} = 2 S_i \left[ \sum_{i' \neq j}^{q} J_{ii'} S_{i'} - \sum_{j' \neq i}^{q} J_{jj'} S_{j'} \right]. \qquad (2)$$

Here, $q$ denotes the coordination number of a site, $i'$ and $j'$ represent the nearest neighbor sites of $i$ and $j$ sites, respectively. The spin-exchange is then accepted or rejected with the standard Metropolis acceptance probability:[38,39]

$$P = \begin{cases} e^{-\beta \Delta H_{ij}}, & \Delta H_{ij} > 0 \\ 1, & \Delta H_{ij} \leq 0 \end{cases} ; \qquad (3)$$

where $\beta = 1/k_B T$. The unit of time for all the following analyses is one Monte Carlo step (MCS), which consists of $N = L_x \times L_y$ spin-exchange attempts by using Eq. (3).

The initial configuration of a critical $AB$ mixture has a random distribution of $A$ ($S_i = +1$) and $B$ ($S_i = -1$) atoms in a 1:1 ratio, as illustrated schematically in Fig. 1(a) is corresponding to a homogeneous state of the system at a high temperature ($T \gg T_c$). This implies no interaction ($J_{ij} = 0$) between the spins at high temperatures. Thus, the exchange probability of $nn$ opposite spin pair $P \to 1$. The homogeneous system is then quenched below $T_c$ for the evolution to take place. We probe the effect of BD on evolution morphologies, growth kinetics, and dynamic universality of the system and its sensitivity to shallow and deep quenches at $T = 0.5$, 1.0, and 1.5, in the asymptotic time limit, which is mostly unexplored for the Ising systems.

Shortly, we present the results for various quench depths at three different fractions of disorder, $f_d = 0.02$, $0.05$, and $0.1$ (see the graphical illustration in Figs. 1(d)-(f)) and compare them with the pure case ($f_d = 0.0$).

*__Correlation function and structure factor:__* To characterize the evolution morphology and the length scale, we compute the two-point equal-time correlation function[1,2], which measures the overlap of spin configuration at a distance $\vec{r}$:

$$C(\vec{r},t) = \frac{1}{N}\sum_i [\langle S_i(t) S_{i+\vec{r}}(t)\rangle - \langle S_i(t)\rangle\langle S_{i+\vec{r}}(t)\rangle]. \tag{4}$$

The angular brackets represent the statistical averaging of data. The structure factor, an experimentally more relevant physical parameter to study the domain morphology, is the Fourier transform of $C(\vec{r},t)$:[1,2]

$$S(\vec{k},t) = \sum_{\vec{r}} e^{i\vec{k}\cdot\vec{r}} C(\vec{r},t), \tag{5}$$

where $\vec{k}$ represents the scattering wave vector. When the evolved morphologies are isotropic, the correlation function and the structure factor statistics can be improved by spherical averaging. The corresponding quantities are denoted as $C(r,t)$ and $S(k,t)$, respectively, where $r$ is the separation between two spatial points and $k$ is the magnitude of wave-vector. The correlation function and the structure factor data are spherically average over ten independent runs unless stated otherwise. For the anisotropic morphologies in the system, we compute the component of the structure factor in different directions.

*__Scaling functions and length scale:__* The domain coarsening is a well-established scaling phenomenon characterized by a unique length scale, $\ell(t)$. The dynamical scaling forms of the correlation function and the structure factor are as follows:

$$C(r,t) \sim g[r/\ell(t)], \tag{6}$$
$$S(k,t) \sim \ell(t)^d f[k\ell(t)], \tag{7}$$

where $g(x)$ and $f(p)$ are scaling functions. The characteristic length scale, $\ell(t)$, is estimated from the correlation function as the distance over which it decays to zero or a fraction of its maximum value, $C(0,t) = 1$.[15] We find that the decay of $C(r,t)$ to 0.1 gives a good measure of $\ell(t)$. A few other definitions of length scale are (i) inverse of the first moment of the structure factor[15] and (ii) the first moment of normalized domain-size distribution.[40,41] They differ only by constant multiplicative factors in the scaling regime.[15,40,41] We extracted the asymptotic growth exponent by computing an effective growth exponent as follows:[28,42,43]

$$\phi_{eff} = \log_\alpha \left[ \frac{\ell(\alpha t)}{\ell(t)} \right], \tag{8}$$

where we set the log-base $\alpha = 2$.

## 3. Simulation results and discussion

We quench the system at $t = 0$ MCS from a high-temperature homogeneous phase to a temperature, $T < T_c$, and monitored the coarsening at various MCS. In displaying these results, we focus on (i) probing the effects of different fractions of disorder ($f_d = 0.02$, 0.05, and 0.1) on coarsening morphologies at shallow and deep quench depths ($T < T_c$), and (ii) how $f_d$ and quench depths influence the system's characteristic growth laws and scaling behavior in the late time limit.

### 3.1. **Pure binary mixture:** $f_d = 0.0$

At the outset of our simulation and to keep the rest of the results in proper context, we first illustrate the well-known kinetics for pure case ($\varepsilon_{ij} = 0, f_d = 0.0$) at three quench temperatures in Fig. 2. The evolution morphologies in Fig. 2 are at $t = 4 \times 10^6$ MCS for (a) $T = 0.5$, (b) $T = 1.0$, and (c) $T = 1.5$, respectively. After the quench, the system evolves with the emergence and growth of domains via SD, showing a typical interconnected, bi-continuous morphology. In the two phases, $A$-rich is marked in maroon, and $B$-rich is unmarked. At $T = 0.5$ (deep quench), due to insignificant thermal fluctuations ($k_B T$), coarsening stays in its early stage[1–3] (*transient growth regime*) of demixing even for $t \approx \mathcal{O}(10^6)$ MCS. However, the system shows typical domain morphologies at moderate to shallow quenches: $T = 1.0$ and $T = 1.5$, respectively, within the same period. The morphology at $T = 1.5$ (Fig. 2(c)) seems a little fuzzier due to a higher thermal noise than for the other two lower temperatures shown in Figs. 2(a)-(b). The insets in Figs. 2(a)-(c) illustrate the spatial intensity variation of the structure factor, $S(k_x, k_y)$, demonstrating the isotropic domain evolution. The color bars on the right indicate the range of scattering intensity values for all the temperatures, indicating that the domain evolution for $T = 0.5$ is in its early stage and faster for $T = 1.5$.

To understand the evolved morphologies of a pure $AB$ mixture at different temperatures, we plot the spherically averaged scaled correlation function, $C(r, t)$ vs. scaled distance, $r/\ell(t)$ as displayed in Fig. 2(d). Data sets for $T = 1.0$ (red curve) and $T = 1.5$ (green curve) nicely collapse onto a single curve. However, a slight deviation is observed at larger domain sizes ($r/\ell(t) \in (1,3)$) for $T = 0.5$ (black curve) as domain evolution is still in the transient growth regime. In the inset of Fig. 2(d), we plot the scaled structure factor, $S(k,t)\ell(t)^{-2}$ vs. scaled distance, $k\ell(t)$. For this and the following $S(k,t)\ell(t)^{-2}$ vs. $k\ell(t)$ plots, we have considered the logarithm of data values on both axes unless stated otherwise. The scaled structure factor curves nicely overlap onto a master curve for higher $k$ values, i.e., for smaller domain sizes for all the temperatures. Whereas at $T = 0.5$, $S(k,t)\ell(t)^{-2}$ (denoted by the black circles) deviates from overlapping with the red and green curves for smaller $k$ values, i.e., larger domain sizes. Therefore, on the time scale of our simulation, the scaling functions demonstrate that the Ising system regards dynamical scaling except for a slight deviation for $T = 0.5$ (deep quench) at smaller $k$ values. Nevertheless, the sizeable $k$ region (tail) of the structure factor ($k \to \infty$) follows Porod's law: $S(k,t) \sim k^{-3}$ which results from the scattering of sharp domain interfaces.[44,45]

The time-dependence of average domain size ($\ell(t)$ vs. $t$) in Fig. 2(e) illustrates that the evolution kinetics is faster for $T = 1.0$ (red curve) and 1.5 (green curve) than at $T = 0.5$ (black curve); the former curves follow LS growth law: $\ell(t) \sim t^{1/3}$.[29] However, the black curve shows the growth exponent: $\phi \to 0.16$, much smaller than LS diffusive growth exponent. To clarify this, we plot the growth exponent $\phi_{eff}$ vs. $1/\ell(t)$ at different temperatures in the inset of Fig. 2(e). At $T = 1.0$ and 1.5, the growth exponent $\phi_{eff} \to 1/3$. However, for $T = 0.5$, $\phi_{eff}$ is still far from the LS growth exponent. This further confirms the slower domain evolution at $T = 0.5$; the domain evolution stays in the transient growth regime for most of the simulation time.

**3.2. Binary mixtures with the lower fractions of bond disorder: $f_d = 0.02, 0.05$**

To see the effect of BD, we first examine the evolution snapshots at $t = 4 \times 10^6$ MCS for $f_d = 0.02$ in Fig. 3(a)-(c) quenched at $T = 0.5, 1.0$, and $1.5$, respectively. At deep quench (in Fig. 3(a)), much smaller domains are formed; the lower $f_d$ seems to have negligible influence on the evolution morphology. When the system is moderately (in Fig. 3(b) at $T = 1.0$) or shallow (in Fig. 3(b) at $T = 1.5$) quenched, we find the formation of short and interconnected stripes or fragmented lamellar, respectively, unlike the pure case where a smooth and bi-continuous morphology is obtained (see Figs. 2(b)-(c)). Notice that the morphology is oriented in a particular direction, as exhibited in Fig. 3(b) and Fig. 3(c). The evolution kinetics due to SD is relatively faster when quenched at $T = 1.5$ (see Fig. 2(c)). Therefore, the short and interconnected stripes formed due to disorder at an early stage of separation kinetics merge to form longer stripes that resemble the fragmented lamellar patterns (see Fig. 3(c)).

The direction of the stripe's alignment depends on the fraction of disordered sites for the given system size. The stripes align in the direction of a higher number of disorder sites on the lattice, as displayed in Figs. 1(d)-(f). Recall that $J_{ij} = 0$ when any of the two $nn$ spins belong to disordered sites. Since phase separation kinetics occurs due to the spin exchange, the most probable locations for this to initiate domain evolution would be in the proximities of disordered sites. Therefore, we start noticing the formation of shorter stripes in the direction of a higher number of disordered sites even at early times, leading to more extended stripe patterns at late times.

In Figs. 3(d)-(g), we plot the scaling functions and the effective growth exponents to characterize the effect of lower BD ($f_d = 0.02$) with quench temperature. Fig. 3(d) presents $C(r,t)$ vs. $r/\ell(t)$ for three quench temperatures (indicated by three different symbols) at $t = 4 \times 10^6$ MCS when the system is already in the scaling regime. The corresponding scaled structure factor, $S(k,t)\ell^{-2}$ vs. $k\ell$ plot is demonstrated in Fig. 3(e). The data in Figs. 3(d) and 3(e) display a slight deviation from the scaling at larger $r/\ell(t) \in$ (1,3) after zero-crossing (or smaller $k\ell(t)$) with quench temperature. The extent of deviation from the scaling for $f_d = 0.02$ with quench depths is similar to the one we observed for a pure system (see Fig. 2(d)). Nevertheless, a good scaling is kept for the sizable portion of the curves. The characteristic length, $\ell(t)$ vs. $t$ plot, is displayed in Fig. 3(f) on a logarithmic scale. Similar to a pure case, the rate of domain size evolution is faster for the shallow quench, $T = 1.5$ (green curve), than for the deep quench, $T = 0.5$ (black curve). The system follows LS growth law ($\phi \to 1/3$) for the shallow quenches. In contrast, the black curve shows the growth exponent: $\phi \to 0.1$. Figure 3(g) displays the effective growth exponent $\phi_{eff}$ against $1/\ell(t)$ at different temperatures, further suggests the same.

The evolution snapshots in Fig. 4(a)-(c) exhibit the late time ($t = 4 \times 10^6$ MCS) morphologies for $f_d = 0.05$ quenched at $T = 0.5, 1.0,$ and $1.5$, respectively. Similar to the case with $f_d = 0.02$, the effect of BD at deep quench is negligible, as demonstrated in Fig. 4(a). However, we observe the formation of short and interconnected stripes (in Fig. 4(b)) at moderate quench temperature and longer stripes resembling the fragmented lamellar (in Fig. 4(c)) at a shallow quench temperature. Again, morphology orientation in both the cases is along the higher number of disordered sites, as exhibited in Fig. 1(e) for $f_d = 0.05$. The scaling functions display a more visible deviation from the master curve, as displayed in Figs. 4(d) and 4(e) for different quenches denoted by different symbol types, compared to $f_d = 0.02$ (in Fig. 3). The tail of the structure factor shows a power-law decay: $S(k,t) \sim k^{-3}$ (Porod's tail) due to scattering from sharp interfaces. The average domain size illustrates the usual LS power-law growth: $\ell(t) \sim t^{1/3}$ for higher temperatures as depicted in Figs. 4(f) and 4(g) with red (at $T = 1.0$) and green (at $T = 1.5$) symbols, respectively. However, the growth at $T = 0.5$ is still the slowest and remains in the transient regime with a growth exponent: $\phi_{eff} \to 0.05$ (see the inset in Fig. 4(g)).

Interestingly, at $T = 0.5$, the growth exponent, $\phi_{eff}$ is getting smaller with the increasing $f_d = 0.0 \to 0.05$ (see insets in Figs. 2(e), 3(g), and 4(g)). The reason could be that the small domain structures evolved at deep quench temperature. Since the system's thermal energy ($k_B T$) is low at $T = 0.5$, thus increasing BD ($f_d = 0.0 \to 0.05$), instead of overcoming domain evolution out of the transient growth stage, begins to melt the morphologies. Hence, the smaller effective growth exponent. But, on the other hand, the domain growth quickly crossover into the diffusive growth regime ($\phi_{eff} \to 1/3$) for moderate and shallow quenches for $f_d = 0.02 \to 0.05$. The green curves in Figs. 2(e), 3(f), and 4(f), respectively, show that the average domain size is more prominent at shallow quench temperature. Although for $f_d = 0.05$, $\phi_{eff}$ crosses over the diffusive growth exponent (1/3) to a relatively little higher value at late times. The reason could be that a higher fraction of BD gradually modifies the domain interfaces. Hence, along with the phase separation dynamics, the random motion (diffusion dynamics) of Ising spins set in at the domain boundaries where the average spin displacement is proportional to $t^{1/2}$.[39] Thus, a crossover to higher $\phi_{eff}$ value at late times is observed.

The morphologies at shallow quench temperature illustrate long stripes (fragmented lamellar) shown in Fig. 3(c) and 4(c) at $t = 4 \times 10^6$ MCS. Although, the spherically averaged structure factor, $S(k,t)$, displayed in Figs. 3(e) and 4(e) successfully demonstrated the system's scaling behavior, not the induced anisotropy caused by stripe morphology. Therefore, we plot $S(k_x, k_y)$ vs. $k_x$ along the lattice diagonals in Figs. 5(a) and 5(b) for $f_d = 0.02$ and $f_d = 0.05$, respectively, to illustrate the same. The black and red curves show the diagonal and cross diagonal structure factors. We average the data over fifty ensembles. The nonoverlapping of $S(k_x, k_y)$ vs. $k_x$ data confirms the presence of structural anisotropy in the system. Notice the interchange of black and red $S(k_x, k_y)$ curves for $f_d = 0.02$ in Fig. 5(a) and $f_d = 0.05$ in Fig. 5(b), they demonstrate the change in stripe's alignment. The corresponding plots of spatial variation of scattering intensity in Figs. 5(c) and 5(d) further verify the change in stripes' orientation.

In addition, we also observe the similar behavior of $S(k_x, k_y)$ vs. $k_x$ curves at the moderate quench ($T = 1.0$) for $f_d = 0.02$ and $f_d = 0.05$, as shown by the green and blue curves in Figs. 6(a) and 6(b), respectively. However, $S(k_x, k_y)$ vs. $k_x$ plots for a deep quench ($T = 0.5$) illustrate an excellent data overlap for the lower fractions of BD shown by the black and red symbols in Fig. 6, thus sustaining the system's isotropy. Overall, the lower $f_d$ values have a negligible effect on the evolving morphologies and the scaling functions at deep quenching. In comparison, shallow quenching at the same fractions of BD considerably influences the morphologies, scaling functions, and length scale of the phase separating system.

### 3.3. Binary mixture with a higher fraction of bond disorder: $f_d = 0.1$

The evolution morphologies for $f_d = 0.1$ is displayed at $T = 0.5$ in Fig. 7(a), $T = 1.0$ in Fig. 7(b), and $T = 1.5$ in Fig. 7(c) for $4 \times 10^6$ MCS. Recall the slow domain evolution at $T = 0.5$ for $f_d = 0.02$ and $f_d = 0.05$, where the system's isotropy was preserved. Herein, at $T = 0.5$ and for a higher fraction of disorder $f_d = 0.1$, the evolving domains begin to form stripes much earlier that evolve into a fragmented lamellar with time, which further evolves to form perfect lamellar patterns at a late time, $t = 4 \times 10^6$ MCS. Thus, when we deep quench the system, a higher fraction of BD seems sufficient to bring the domain evolution out of the slower transient regime. Hence, domains begin to evolve into stripe morphologies. A few tiny random domains of $A$-type (marked in maroon) are seen in the $B$-type phases (unmarked) displayed in Figs. 7(a)-(c). The fuzziness in the system is due to large thermal fluctuation (noise) at $f_d = 0.1$. Note that we observe only fragmented lamellar structures at shallow quenches for lower $f_d$ values even at late times. However, for $f_d = 0.1$, we get similar morphologies much earlier, which form the perfect lamellar patterns at late times. Therefore, when a high fraction of BD is introduced into the system, perfect anisotropic structures develop at late times for all the quench depths discussed here.

Next, we compare the scaling functions and length scales for $f_d = 0.1$ in the asymptotic time limit, $t = 4 \times 10^6$ MCS for three different quenches, $T = 0.5, 1.0,$ and $1.5$ in Fig. 7(d)-(g). The data sets for $C(r, t)$ vs. $r/\ell(t)$ and $S(k, t)\ell(t)^{-2}$ vs. $k\ell(t)$ do not overlap, as demonstrated in Figs. 7(d) and 7(e) by the black ($T = 0.5$), red ($T = 1.0$), and green ($T = 1.5$) symbols. The deviation from the scaling ensures that the evolved morphologies do not belong to the same universality class. The solid black line in Fig. 7(e) with slope $= -3$ shows the structure factor tail deviation from Porod's law for $k \to \infty$. Porod's law results from the scattering off sharp domain interfaces of small domain structures. Due to thermal noise at a higher BD, the sharp domain interfaces become fuzzier, hence, the deviation from the well-known Porod's law. The oscillations in $S(k,t)\ell(t)^{-2}$ vs. $k\ell(t))$ curves confirm the periodicity in the system due to lamellar morphology. We observed that with increasing quench temperature, the main peak of the structure factor gets narrower for lower $k$, and the oscillatory behavior of $S(k,t)$ is enhanced for larger $k$; this characterizes the periodicity developed in the system due to lamellar patterns at a late time.

The characteristic length scale follows LS growth law ($\phi \sim 1/3$) at early times for all the quench temperatures indicated in Fig. 7(f). During the intermediate time, the growth law gradually crossover to diffusion kinetics ($\phi \sim 1/2$) due to the random motion of spins at domain interfaces caused by higher $f_d$. Beyond $t > 10^6$ MCS, the domain growth freezes to a finite value at moderate and shallow quenches (see the red and green curves at $T = 1.0$ and 1.5, respectively). The system attains its equilibrium lamellar morphology earlier for higher quench as the growth kinetics is relatively faster for higher temperatures. We plot $S(k_x, k_y)$ vs. $k_x$ along the diagonals in Fig. 7(g) to compare the anisotropy in the system due to lamellar morphology obtained at deep and shallow quenches. For $T = 0.5$, the black and red curves, and for $T = 1.5$, the green and blue curves show $S(k_x, k_y)$ vs. $k_x$ along the diagonal and cross-diagonal of the lattice, respectively. The structure factor peak retains a much higher amplitude when computed diagonal (normal to the stripes) than cross-diagonal (along the strips). Thus the nonoverlapping of $S(k_x, k_y)$ vs. $k_x$ curves demonstrate the presence of anisotropy in the system. However, a relatively higher peak strength of $S(k_x, k_y)$ at $T = 1.5$ further suggests that evolved morphologies are more anisotropic at a shallow quench than the morphologies at $T = 0.5$.

Our results in Figs. 8(a) and 8(b) for the late time $t = 4 \times 10^6$ MCS) suggest that for the lower fractions of BD at $T = 0.5$, the scaling functions (the spherically averaged correlation function and the structure factor) collapse almost neatly onto the scaling function of a pure system ($f_d = 0.0$), indicated by the black, red, and green symbols. Therefore, the system quenched at $T = 0.5$ with $f_d = 0.0$, 0.02, and 0.05 belong to the same dynamical universality class. In other words, the morphologies appear nearly similar to the pure system when evolved at $T = 0.5$ within a lower $f_d$. The symmetry in the scattering intensity variation for $f_d = 0.02$ (in Fig. 8(c)) and $f_d = 0.05$ (in Fig. 8(d)), further justifies the isotropy of evolved $A$ and $B$ phases within a minor deviation from the pure case (in Fig. 2(a)). Hence, they belong to the same dynamical universality class. However, for $f_d = 0.1$, the morphology of $A$ and $B$ phases transformed into lamellar patterns. Hence, the deviation from the dynamical scaling function (indicated by the blue symbols in Figs. 8(a)-(b)) shows that system does not belong to the same universality class. The changing symmetry patterns in the scattering intensity variation in Fig. 8(e) manifest the same. Recall that when the system was quenched at $T = 1.0$ or $1.5$, the anisotropy was induced even at lower fractions of BD in the form of short/long stripe patterns. Whereas, when BD is increased up to $f_d = 0.1$, the strips transform into a broken lamellar pattern which becomes a perfect lamellar at late times for all the quench temperatures.

## 4. Summary and conclusion

We studied the effect of different quench temperatures, $T = 0.5, 1.0,$ and $1.5$ $(T < T_c)$ on phase separation kinetics of a critical binary $(AB)$ mixture under the influence of different fractions of disorder, $f_d = 0.0$ (pure case), $0.02$, $0.05$, and $0.1$, introduced in a regular manner. The separation kinetics was modeled using the conserved (Kawasaki) spin-exchange dynamics on the $2d$ kinetic Ising model by utilizing the Monte Carlo (MC) simulation technique. In particular, we explored how the deep $(T = 0.5)$ and shallow $(T = 1.5)$ quenches at various $f_d$ influence the system's morphology, characteristic growth laws, and scaling behavior in the late time limit.

When a homogeneous binary mixture was deep quenched, the influence of lower $f_d$ on the segregation kinetics was almost negligible; the system showed nearly perfect dynamical scaling with the pure case, displayed by the scaled correlation function and the structure factor. The domain evolution stayed within a transient growth regime even at an asymptotic time limit. The growth exponent, $\phi \sim 0.16$ for the pure case, decreased even further; $\phi \sim 0.05$ for $f_d = 0.05$ due to some randomness at domain interfaces caused by the disorder. Thus, the growth exponent for a binary mixture, deep quenched at $T = 0.5$ with $f_d = 0.02$ and $0.05$ were much smaller than the usual *Lifshitz-Slyozov* (LS) growth exponent, $\phi \sim 1/3$.

On the other hand, when the system was shallow quenched $(T = 1.5)$, the effect of lower fractions of disorder was quite apparent in the evolution kinetics. Therefore, we have also considered the quench at a moderate temperature, $T = 1.0$ for a better comparison. A pure system displayed the usual evolution kinetics at the temperatures mentioned above.

However, the system evolved to form short and interconnected stripe patterns for the lower $f_d$, quenched at $T = 1.0$; longer and fragmented lamellar morphology was observed when quenched at $T = 1.5$. These morphologies were oriented particularly toward a higher number of disorder sites. Furthermore, the system evolved into anisotropic morphologies; hence the deviation from dynamic scaling was observed that became more noticeable with the increase of $f_d$ at shallow quench temperature. However, the growth law was essentially consistent with LS power-law growth with a deviation to a slightly higher growth exponent in the asymptotic time limit.

At a high fraction of disorder ($f_d = 0.1$), we observed the formation of long fragmented strips at early times for all the quench temperatures studied here. These stripe patterns gradually transformed into a perfect lamellar morphology with different periodicity. Therefore, we observed a significant deviation in the dynamical scaling functions at different quench temperatures at the asymptotic time limit. The domain growth was initially consistent with the LS growth law: $\phi \sim 1/3$, which gradually crossed over to the diffusion dynamics: $\phi \sim 1/2$ during the intermediate time. However, on the time scale of our simulation, the domain growth was frozen to a finite size when the system formed its equilibrium (stable) lamellar morphology. We noted that the domain growth at $T = 1.5$ reached saturation earlier than at $T = 1.0$. Nevertheless, we have not accessed the saturation in the length scale for $T = 0.5$; however, there is a possibility to access the same at a much later time than investigated here.

The pattern formation in the disordered system is of great technological importance. Our model can be integrated to understand a wide variety of physical phenomena such as the structural evolution in the biological system (e.g., the iridescent color patterns of bird feathers observed due to refraction of incident light from the phase-separated frozen nanostructures), self-organizing spatial patterns in ecological systems (e.g., mussel beds, etc.). Furthermore, the phase separation kinetics in multiphase fluid and mineral exsolution can be easily explained.

Finally, our simulation results provide a general framework for the experiments on domain growth in kinetic Ising systems with the bond disorder. In the future, we aim to address its effect and quench depths for the Ising systems in $3d$ where similar experiments have not yet been performed. Thus, we hope our simulation results offer essential guidelines for future studies.

**Conflicts of interest**

There are no conflicts of interest to declare.

**Author Contributions**

Conceptualization, A.S., and S.S.; Methodology, S.S.; Validation, A.S.; Formal analysis and Investigation, S.S., and A.S.; Resources and Data curation, S.S.; Writing-original draft preparation, S.S.; Writing-review and editing, A.S.; Visualization, S.S.; Supervision, A.S; Funding acquisition, A.S. All authors have read and agreed to the published version of the manuscript.

**Acknowledgments**

S. S. would like to acknowledge IIT (BHU) for financial support. In addition, A. S. acknowledges the financial support from SERB Grant No. ECR/2017/002529 by the Department of Science and Technology, New Delhi, India.

**References**

1. S. Puri and V. K. Wadhawan, *Kinetics of Phase Transitions* (CRC Press, Boca Raton, FL, 2009).

2. A. J. Bray, Advances in Physics **43**, 357 (1994).

3. A. Onuki, *Phase Transition Dynamics* (Cambridge University Press, 2002).

4. K. Binder and P. Fratzl, in *Materials Science and Technology* (Wiley-VCH Verlag GmbH & Co. KGaA, Weinheim, Germany, 2013).

5. R. A. L. Jones, *Soft Condensed Matter* (Oxford University Press, Oxford, 2008).

6. H. Ikeda, Y. Endoh, and S. Itoh, Physical Review Letters **64**, 1266 (1990).

7. V. Likodimos, M. Labardi, and M. Allegrini, Physical Review B **61**, 14440 (2000).

8. V. Likodimos, M. Labardi, X. K. Orlik, L. Pardi, M. Allegrini, S. Emonin, and O. Marti, Physical Review B **63**, 064104 (2001).

9. A. J. Bray and K. Humayun, Journal of Physics A: Mathematical and General **24**, L1185 (1991).

10. D. J. Srolovitz and G. S. Grest, Physical Review B **32**, 3021 (1985).

11. G. S. Grest and D. J. Srolovitz, Physical Review B **32**, 3014 (1985).

12. J. H. Oh and D.-I. Choi, Physical Review B **33**, 3448 (1986).

13. D. Chowdhury, M. Grant, and J. D. Gunton, Physical Review B **35**, 6792 (1987).

14. M. F. Gyure, S. T. Harrington, R. Strilka, and H. E. Stanley, Physical Review E **52**, 4632 (1995).

15. A. Singh, A. Mukherjee, H. M. Vermeulen, G. T. Barkema, and S. Puri, Journal of Chemical Physics **134**, 044910 (2011).

16. A. Singh, A. Singh, and A. Chakraborti, Journal of Chemical Physics **147**, 124902 (2017).

17. S. Majumder, S. K. Das, and W. Janke, Physical Review E **98**, 042142 (2018).

18. R. Agrawal, M. Kumar, and S. Puri, Physical Review E **104**, 044123 (2021).

19. M. Tateno and H. Tanaka, Nature Communications 2021 12:1 **12**, 1 (2021).

20. H. Manzanarez, J. P. Mericq, P. Guenoun, and D. Bouyer, Journal of Membrane Science **620**, 118941 (2021).

21. A. Singh, Bulletin of Materials Science 2020 43:1 **43**, 1 (2020).

22. K. Takae and H. Tanaka, Proc Natl Acad Sci U S A **117**, 4471 (2020).

23. J. Fan, T. Han, and M. Haataja, The Journal of Chemical Physics **133**, 235101 (2010).

24. R. Paul, S. Puri, and H. Rieger, Physical Review E **71**, 061109 (2005).

25. R. Paul, S. Puri, and H. Rieger, Europhysics Letters (EPL) **68**, 881 (2004).

26. S. Puri and N. Parekh, Journal of Physics A: Mathematical and General **25**, 4127 (1992).

27. S. Puri, D. Chowdhury, and N. Parekh, Journal of Physics A: Mathematical and General **24**, L1087 (1991).

28. D. A. Huse and C. L. Henley, Physical Review Letters **54**, 2708 (1985).

29. I. M. Lifshitz and V. V. Slyozov, Journal of Physics and Chemistry of Solids **19**, 35 (1961).

30. A. Singh, S. Puri, and C. Dasgupta, The Journal of Chemical Physics **140**, 244906 (2014).

31. T. Koga and K. Kawasaki, Physical Review A **44**, R817 (1991).

32. V. M. Kendon, J.-C. Desplat, P. Bladon, and M. E. Cates, Physical Review Letters **83**, 576 (1999).

33. T. Nattermann and J. Villain, Phase Transitions **11**, 5 (1988).

34. A. J. Bray and M. A. Moore, Journal of Physics C: Solid State Physics **18**, L927 (1985).

35. M. Kumar, V. Banerjee, and S. Puri, EPL **117**, (2017).

36. V. Likhosherstov, Y. Maximov, M. Chertkov, L. Vidmar, M. Rigol, I. Balog, G. Tarjus, M. Tissier, A. Bupathy, M. Kumar, V. Banerjee, and S. Puri, Journal of Physics: Conference Series **905**, 012025 (2017).

37. L. Onsager, Physical Review **65**, 117 (1944).

38. M. E. J. Newman and G. T. Barkema, *Monte Carlo Methods in Statistical Physics* (Oxford University Press, Oxford, UK, 1999).

39. D. P. Landau and K. Binder, *A Guide to Monte Carlo Simulations in Statistical Physics*, 4th edition (Cambridge University Press, Cambridge, Massachusetts, 2014).

40. S. Puri and H. L. Frisch, International Journal of Modern Physics B **12**, 1623 (1998).

41. Y. Oono and S. Puri, Physical Review Letters **58**, 836 (1987).

42. A. Chakrabarti, R. Toral, and J. D. Gunton, Physical Review B **39**, 4386 (1989).

43. S. Majumder and S. K. Das, Physical Chemistry Chemical Physics **15**, 13209 (2013).

44. G. Porod, in *Small Angle X-Ray Scattering*, edited by O. Glatter and O. Kratky (Academic Press, New York, 1982).

45. Y. Oono and S. Puri, Modern Physics Letters B **02**, 861 (1988).

**Figures**

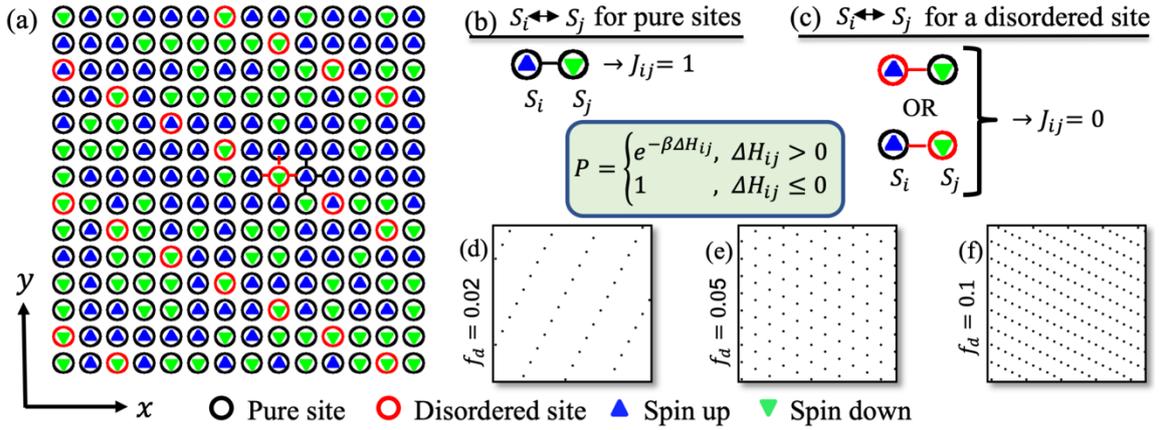

Figure 1: (a) Schematic representation of 2d Ising model with disorder sites on a square lattice. The pure (black circle) and disordered sites (red circle) are tagged with a random distribution of up (blue triangle) and down (green triangle) spins. The standard Kawasaki spin-exchange kinetics probability and the exchange interaction ($J_{ij}$) for the neighboring spins are displayed in (b) when two nearest-neighbor ($nn$) spins are at pure sites, and (c) when at least one $nn$ spins are at the disorder site. (d-f) Display a section of disordered sites on a 2d square lattice of size $N = L^2$ where $L = 512$ for $f_d = 0.02, 0.05$, and $0.1$.

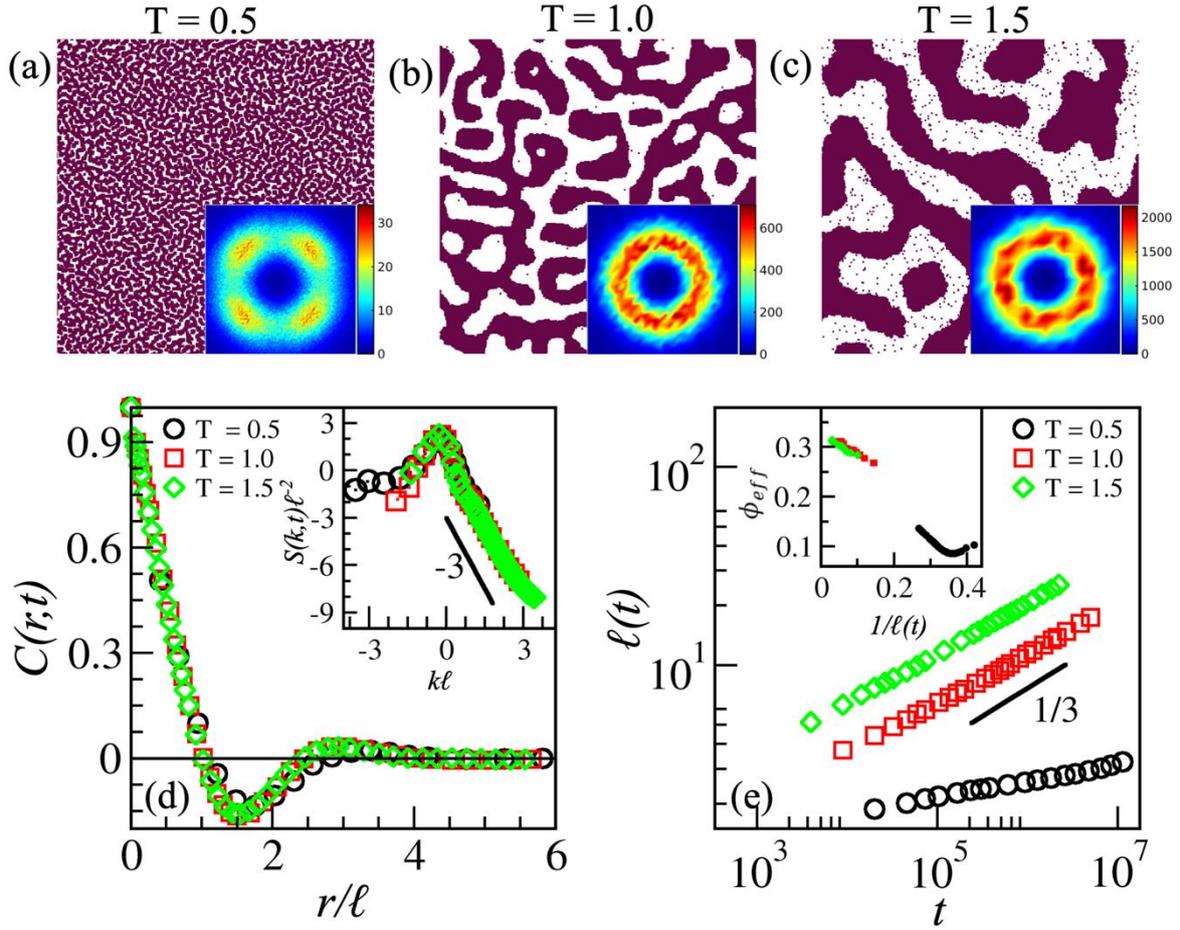

Figure 2: (a) Evolution snapshots for a pure ($f_d = 0.0$) critical binary ($AB$) mixture at $t = 4 \times 10^6$ MCS for (a) $T = 0.5$, (b) $T = 1.0$, and (c) $T = 1.5$. The insets in (a-c) illustrate the spatial intensity variation of $S(k_x, k_y)$. (d) Data sets for the scaling plot of $C(r,t)$ vs. $r/\ell(t)$ at $T = 0.5, 1.0,$ and $1.5$ collapse nicely onto a single curve. The inset shows the scaling plot of $S(k,t)\ell(t)^{-2}$ vs. $k\ell(t)$. The structure factor tail follows Porod's law, $S(k,t) \sim k^{-3}$ for $k \to \infty$. (e) Displays the log-log plot of characteristic length scale, $\ell(t)$ vs. $t$ for the evolution is shown in (a-c). The solid black line shows the expected growth exponent, $\phi = 1/3$ for the pure case. Inset plots show the variation of effective growth exponent, $\phi_{eff}$ as a function of $1/\ell(t)$.

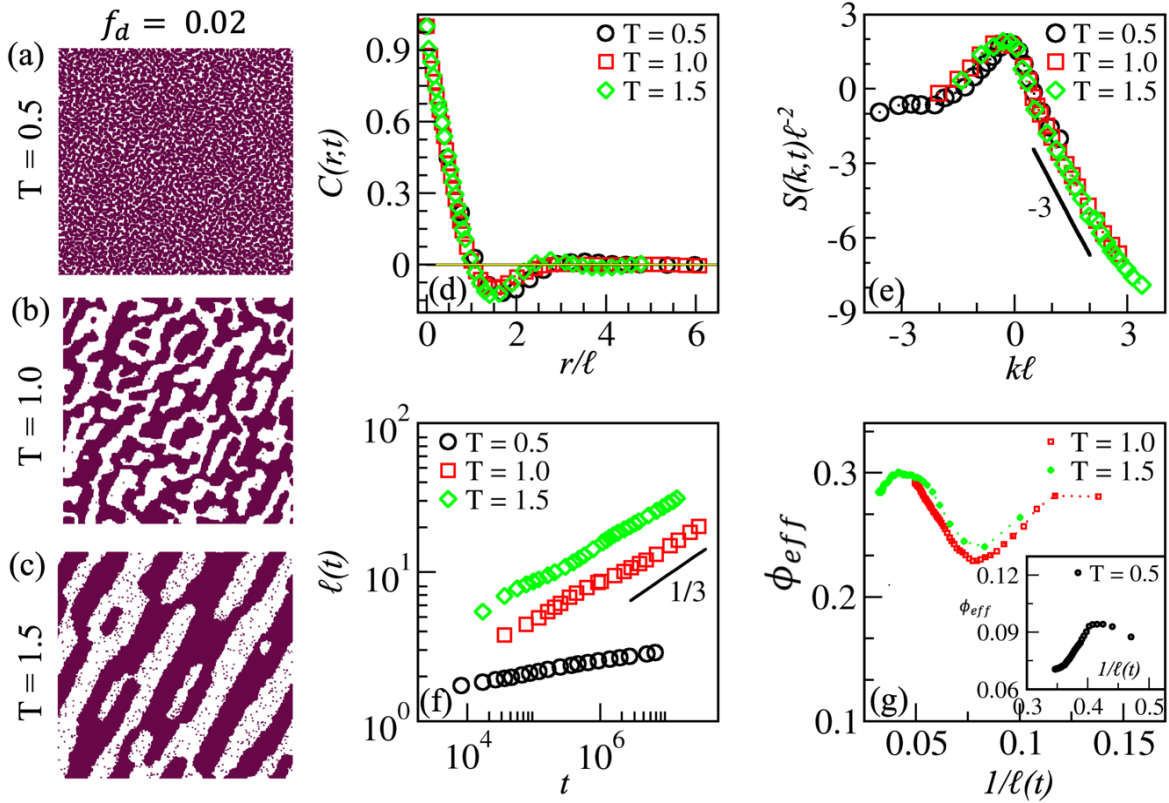

Figure 3: (a-c) Evolution snapshots for $f_d = 0.02$ at $t = 4 \times 10^6$ MCS for $T = 0.5, 1.0$ and $1.5$, respectively. The scaling plot of $C(r,t)$ vs. $r/\ell(t)$ in (d) and $S(k,t)\ell(t)^{-2}$ vs. $k\ell(t)$ in (e) for $f_d = 0.02$ at three quench temperatures are indicated by the different symbols. (f) The log-log plot of the length scale, $\ell(t)$ vs. $t$, and (g) the effective growth exponent, $\phi_{eff}$ vs. $1/\ell(t)$ for the length scale shown in (f) at $T = 1.0$ and $1.5$. The inset in (g) shows $\phi_{eff}$ for $T = 0.5$.

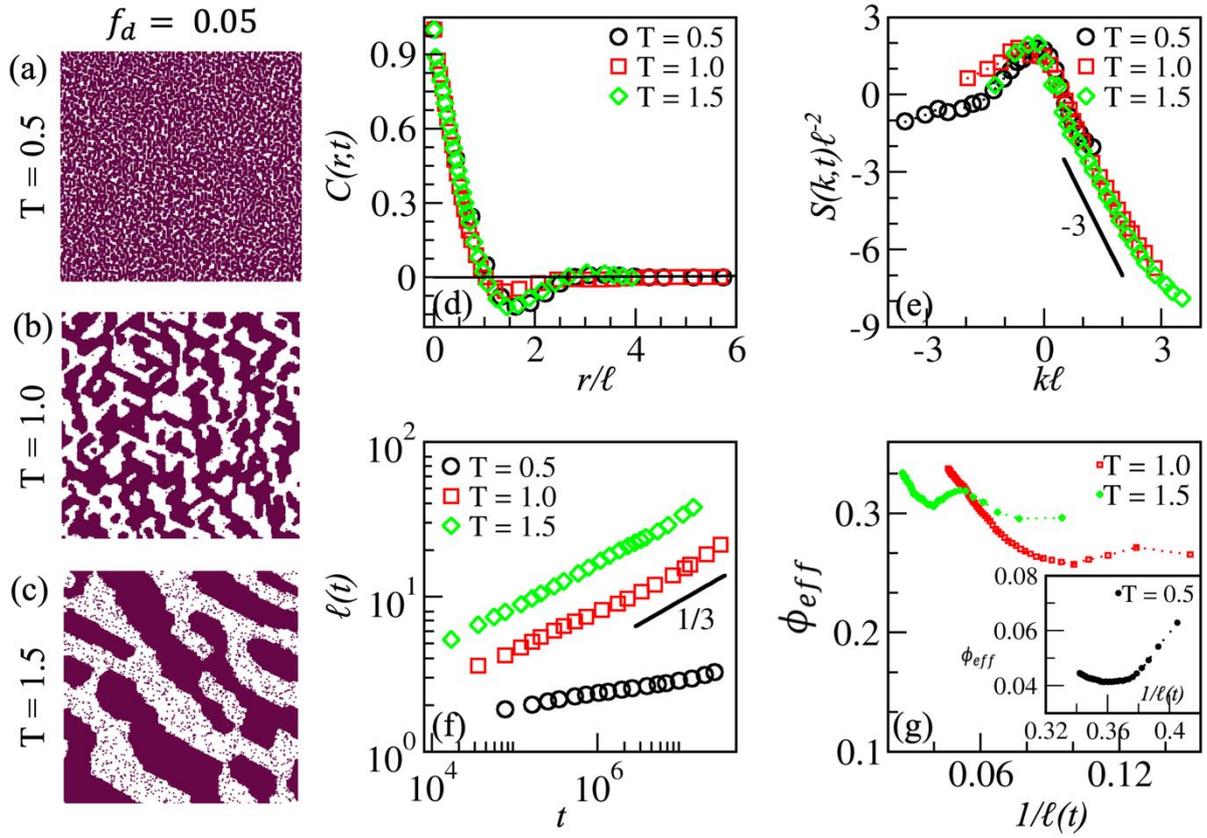

Figure 4: (a-c) Evolution snapshots for $f_d = 0.05$ at $t = 4 \times 10^6$ MCS for different quenches. The corresponding scaling plots of $C(r,t)$ vs. $r/\ell(t)$ in (d) and $S(k,t)\ell(t)^{-2}$ vs. $k\ell(t)$ in (e) at different quenches shown by the various symbol types. (f) Log-log plot of length scale, $\ell(t)$ vs. $t$. (g) The effective growth exponent, $\phi_{eff}$ vs. $1/\ell(t)$ is related to the patterns in (a-c). The inset in (g) shows $\phi_{eff}$ vs. $1/\ell(t)$ for $T = 0.5$.

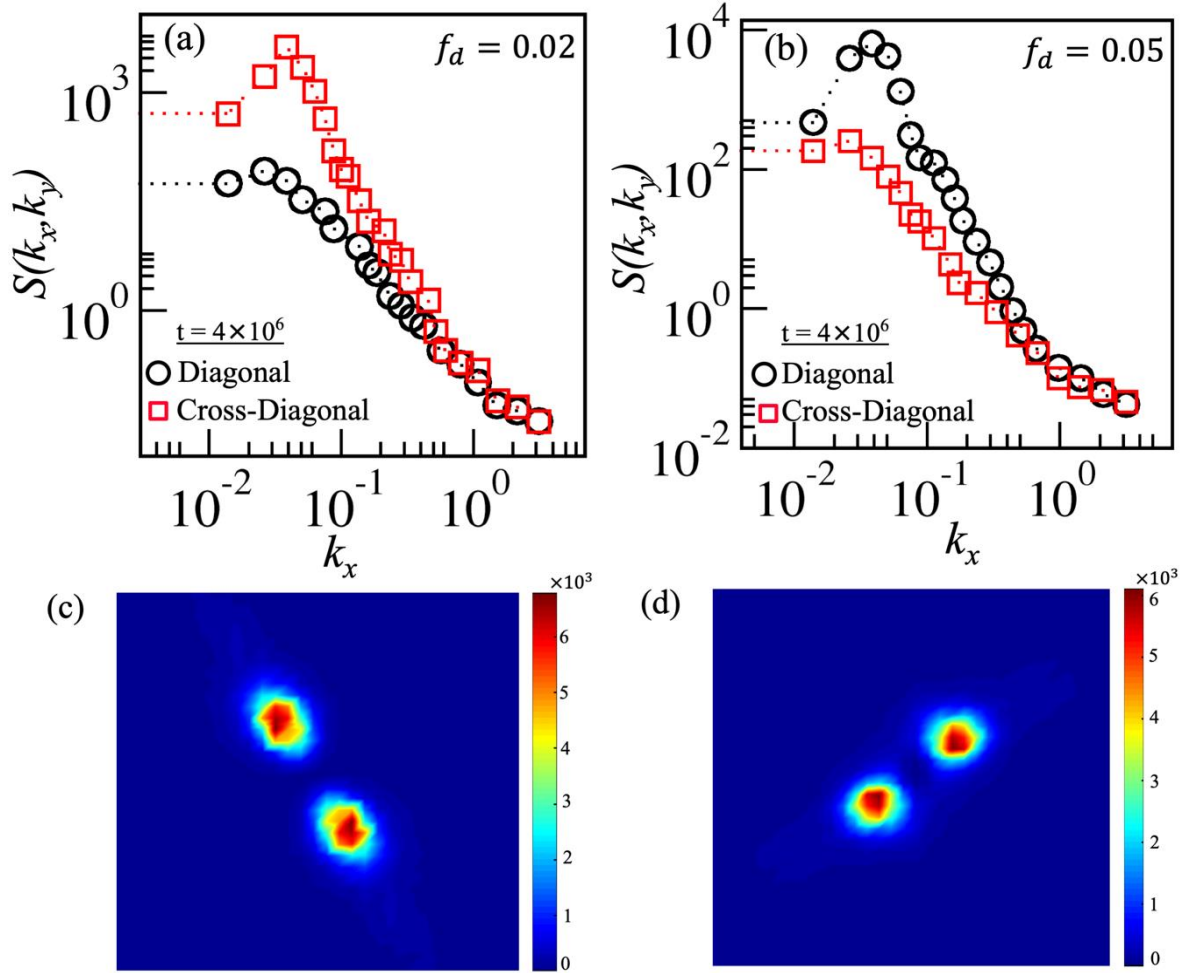

Figure 5: Plots of $S(k_x, k_y)$ along the lattice diagonals at $t = 4 \times 10^6$ MCS for (a) $f_d = 0.02$, and (b) $f_d = 0.05$, respectively at $T = 1.5$. (c-d) shows the corresponding spatial intensity variation of the structure factor, $S(k_x, k_y)$ depicting the orientation and anisotropy in the system.

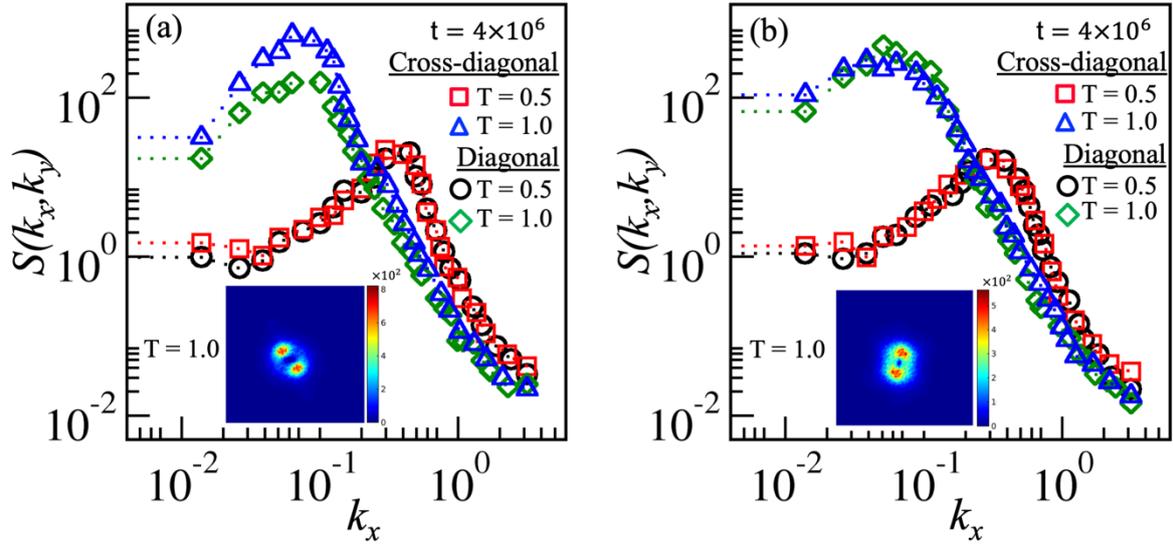

Figure 6: (a) Plot of $S(k_x, k_y)$ vs. $k_x$ across the lattice diagonals for $f_d = 0.02$ at $T = 0.5$ (black and red curves) and $T = 1.0$ (green and blue curves) at $t = 4 \times 10^6$ MCS. (b) Shows the same as in (a) for $f_d = 0.05$. The insets (a) and (b) display the spatial scattering intensity variation at $T = 1.0$ for $f_d = 0.02$, and 0.05.

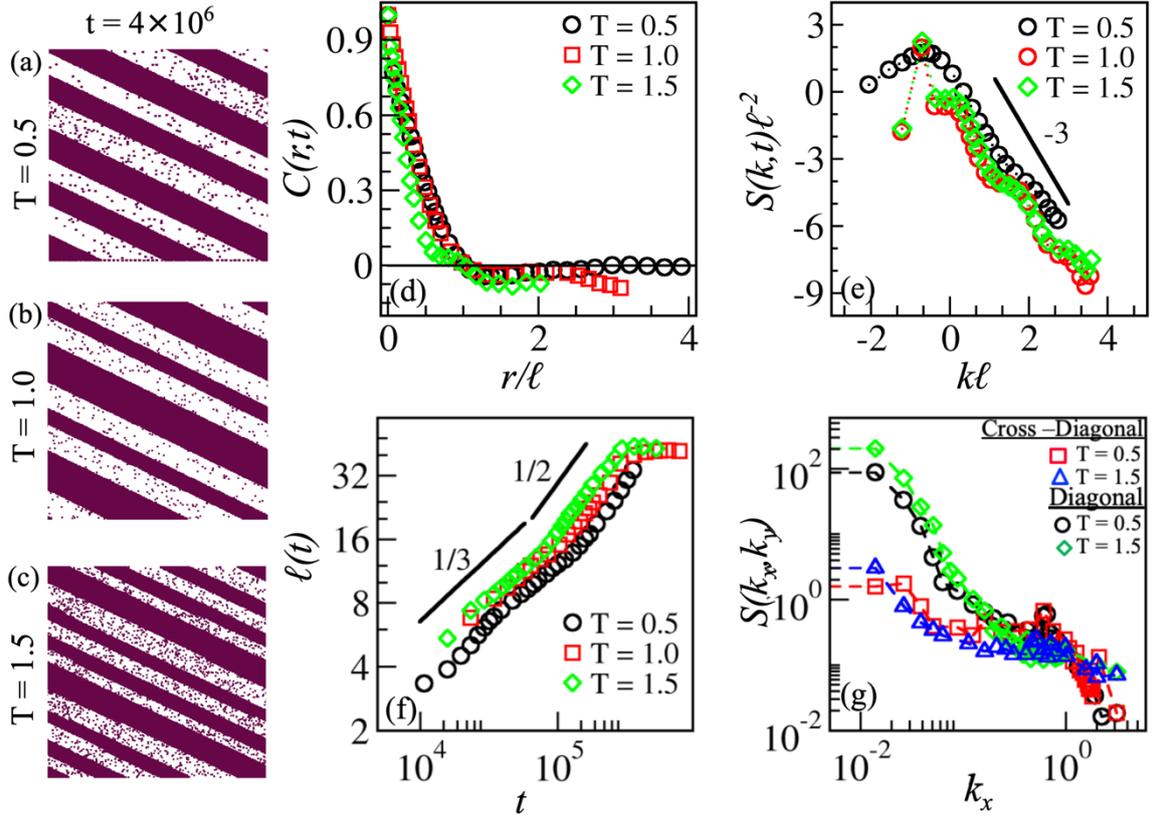

Figure 7: (a-c) Evolution snapshots for $T = 0.5$, $1.0$, and $1.5$, respectively, at $f_d = 0.1$ in the asymptotic limit. (d) The plot of $C(r,t)$ vs. $r/\ell(t)$ for evolutions in (a-c) at $T = 0.5$ (black symbol), $T = 1.0$ (red symbol), and $T = 1.5$ (green symbol). (e) The plot of $S(k,t)\ell(t)^{-2}$ vs. $k\ell(t)$ corresponding to the data sets in (d). (f) Log-log plot of length scale, $\ell(t)$ vs. $t$. (g) The comparison of $S(k_x, k_y)$ along the lattice diagonals for morphologies at $T = 0.5$ and $T = 1.5$.

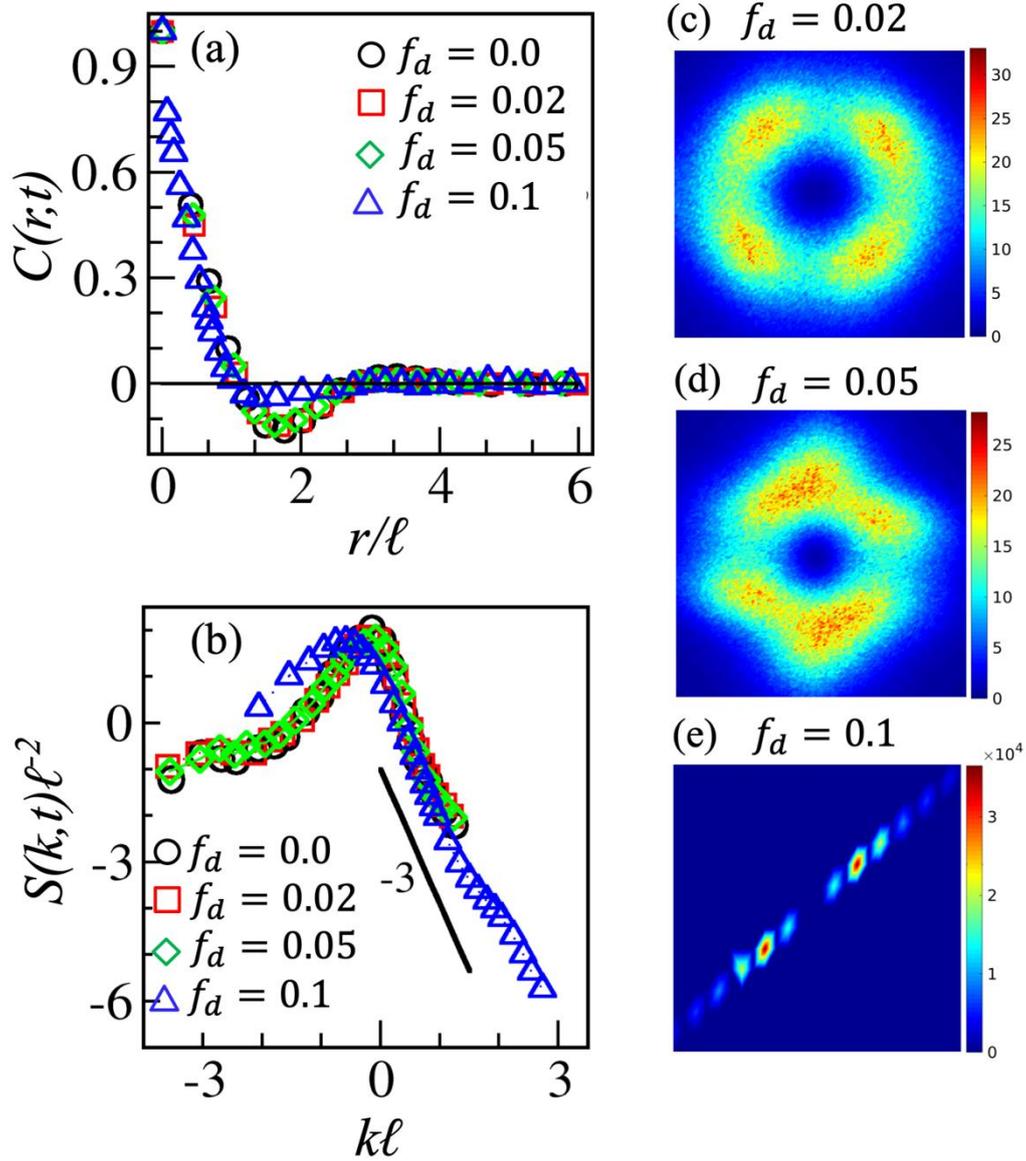

Figure 8: (a) Plot of $C(r,t)$ vs. $r/\ell(t)$ for various fractions of disorder, $f_d = 0.0$ (black curve), $0.02$ (red curve), $0.05$ (green curve), and $0.1$ (blue curve) at $T = 0.5$ for $t = 4 \times 10^6$ MCS. (b) Plot of $S(k,t)\ell(t)^{-2}$ vs. $k\ell(t)$ for the data sets in (a). The spatial variation of scattering intensity at $T = 0.5$ for $f_d = 0.02$ in (c), $f_d = 0.05$ in (d), and $f_d = 0.1$ in (e).

# Supplementary Information

**Phase separation kinetics of binary mixture in the influence of bond disorder: Sensitivity to quench temperature**

Samiksha Shrivastava and Awaneesh Singh*

Department of Physics, Indian Institute of Technology (BHU), Varanasi-221005, India.

## 1. The sensitivity of results to the system size

In the main text, we present all the results for a $2d$ square lattice of size $N = L^2$ where $L = 512$ with periodic boundary conditions in all the directions for three different percentages of BD. In the following, we performed a few more experiments on two smaller system sizes ($L = 128, 256$) to demonstrate that our results are independent of the system size. Note that the number of lattice sites, $N$ changes with the system size $L$. The number of disordered sites, $N_d$ also revises with changing fractions of $N$ for given system size. Thus, the arrangement of a higher number of disorder sites in a particular direction also modifies. Hence, the stripes' orientation can also alter accordingly. Nevertheless, the domain morphologies are similar for different system sizes within the thermodynamic limit. Figures S1 and S2 demonstrate that the evolved stripe patterns are nearly the same for different system sizes, i.e., the stripe's orientation could change but not the patterns.

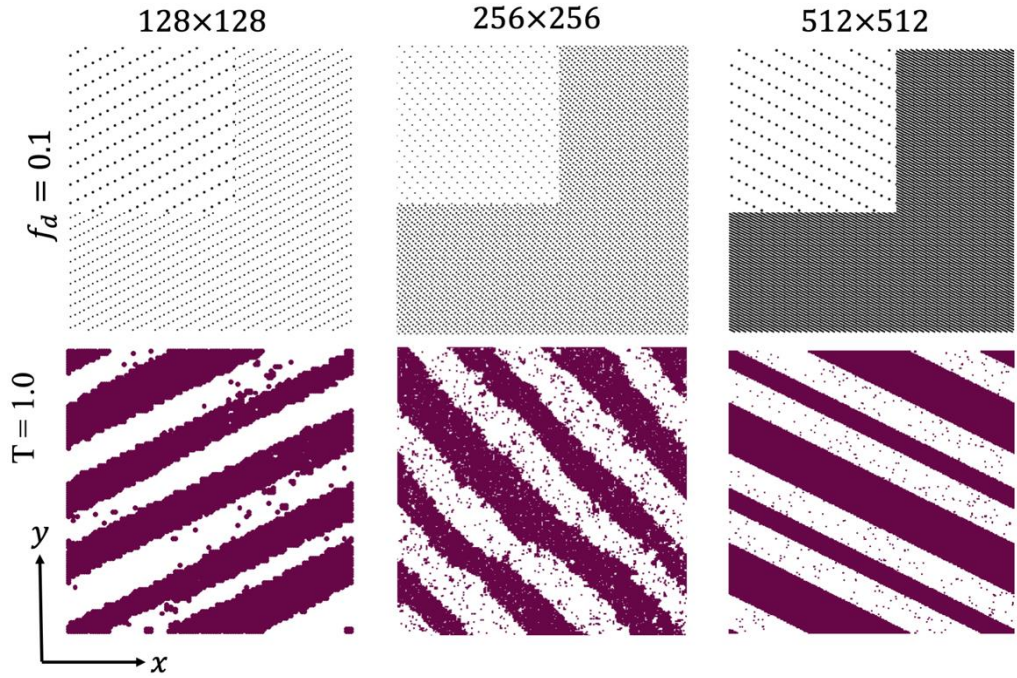

Figure S1: Phase separation for three different system sizes, $N = L^2$ where $L = 128,\ 256$, and $512$, respectively, for a given fraction of disorder, $f_d = 0.1$ at a quenching temperature $T = 1.0$. The top row displays the arrangement of disorder sites, and their top-left corners show the zoomed version of a section of the disorder sites. The bottom row demonstrates the corresponding statistically similar phase-separated morphologies (long stripes or lamellar patterns), oriented in different directions (along with the higher number of disorder sites).

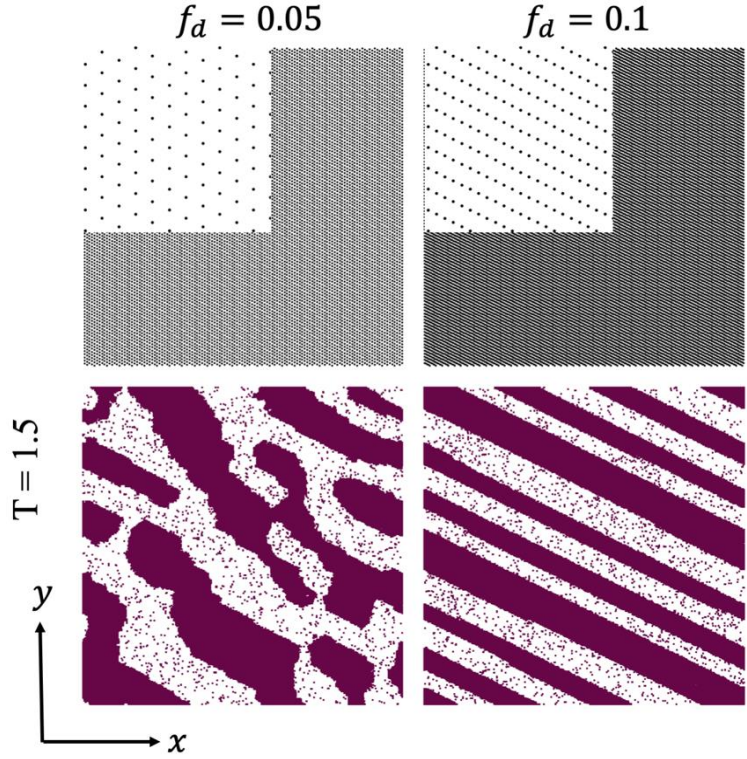

Figure S2: Phase separation for the different fractions of disorder at a fixed system size ($N = L^2$; $L = 512$), quenched at $T = 1.5$. In the top row, we show the distribution of disorder sites $f_d = 0.05$ and $0.1$. The bottom row demonstrates corresponding stripe patterns of different types and orientations (along with the higher number of disorder sites).

2. **Late time comparison of anisotropy at the lower $f_d$ for T = 1.0 and T = 1.5**

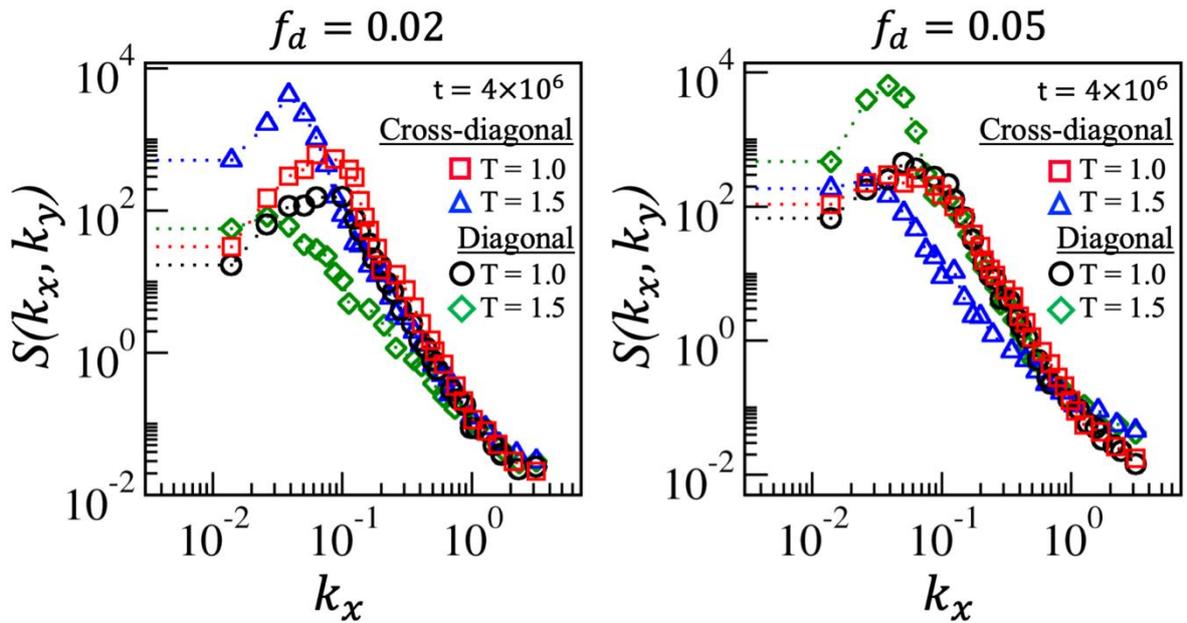

Figure S3: We compare the structure factors without spherical averaging, $S(k_x, k_y)$ vs. $k_x$ along the lattice diagonals at T = 1.5 and T = 1.0 for $f_d$ = 0.02 and $f_d$ = 0.05, respectively. The nonoverlapping of curves confirms the presence of structural anisotropy in the system.